\documentclass[
aps,prd,
12pt,
nopreprintnumbers,
showpacs,
eqsecnum,
nofootinbib
]{revtex4-1}

\usepackage{graphicx}
\usepackage{amssymb}

\begin{document}

\title{Equivalent Hamiltonian approach to quantum cosmology of integrable
models}
\author{Nahomi Kan}\email[]{kan@gifu-nct.ac.jp}
\affiliation{National Institute of Technology, Gifu College,
Motosu-shi, Gifu 501-0495, Japan
}
\author{Masashi Kuniyasu}\email[]{mkuni13@yamaguchi-u.ac.jp}
\author{Kiyoshi Shiraishi}\email[]{shiraish@yamaguchi-u.ac.jp}
\author{Kohjiroh Takimoto}\email[]{i016vb@yamaguchi-u.ac.jp}
\affiliation{
Graduate School of Sciences and Technology for Innovation, Yamaguchi
University, Yamaguchi-shi, Yamaguchi 753--8512, Japan}
\date{\today}

\begin{abstract}
We propose an approach to quantum cosmology of integrable models.
To analyze the models with two dynamical variables, we introduce equivalent
Hamiltonians in reduced phase spaces,
which are obtained with the aid of the Faddeev--Jackiw method. Quantum dynamics of
the models can be studied by using the equivalent Hamiltonians with various
techniques.
\end{abstract}



\maketitle

\section{Introduction}
\label{sec1}

In study of quantum cosmology \cite{HH,Hawking,Halliwell,Kiefer0}, the
probabilistic interpretation of the wave function of the Universe has not been
well-defined, because the Wheeler--DeWitt wave equation, which comes from the
Hamiltonian constraint, is the second order differential equation in terms of the
minisuperspace variables. In other words, one can hardly extract any conserved
probability current from the Wheeler--DeWitt wave function of the Universe.

The problem of the probability distribution is closely related to the
problem of time in quantum cosmology.
If the basic equation is the Schr\"odinger-type equation that is the first-order
differential equation in time, one can obtain the probability
density as the absolute square of the wave function.

Many attempts have been made to define a time variable in quantum cosmology:
some authors found `global time' \cite{Hajicek,Simeone}, and others defined
`extrinsic time' \cite{CF} or `conformal time' \cite{Kuzmichev}, etc..
Especially, various models defined in two dimensional minisuperspace have been
investigated in this context, and one soluble variable in such a model often plays
a role of a standard `clock' in the Universe.
The other approaches utilize
decomposition of the Hamiltonian and the wave function \cite{Mostafazadeh}
or introduce an effective Hamiltonian \cite{HA}.
Even in the former attempt for defining the time, of course,
the Hamiltonian defined in each work is different from the original one in a
certain sense.

In the present paper, we propose another type of
equivalent Hamiltonians for two integrable models.
We adopt a heuristic approach to obtain an equivalent Hamiltonian.
The general idea of treating a constraint system is explained in the following
text in the succeeding section.

The structure of the present paper is as follows.
In Sec.~\ref{sec2}, we specify the first model we treat here, the Liouville scalar
model. Although an equivalent Hamiltonian is introduced by a heuristic
consideration, the idea of this `trick' is briefly addressed here.
Another model, the conformal scalar model is presented in Sec.~\ref{sec3}.
In Sec.~\ref{sec4}, we manage to extract quantum nature of the Liouville scalar
model by using the equivalent Hamiltonians and the method of cumulants.
In Sec.~\ref{sec5}, we treat the wave function of the conformal scalar model.
We examine usefulness of the Gaussian wave packet and the Wigner function in the 
model. The last section is devoted to conclusion and prospect. In Appendix
\ref{AA}, we describe the canonical transform of the Liouville mechanics.

\section{the Liouville scalar model}
\label{sec2}

\subsection{The original system}

The cosmological gravitational model of a scalar field with an exponential
potential has been studied by many authors until recent times (see, for example,
Refs.~\cite{Russo,Neupane,ENO}). We call the model as the Liouville scalar model.
We start with the field theoretic action
\begin{equation}
S=\int d^4x\sqrt{-g}\left[R-\frac{1}{2}(\nabla\Phi)^2-\frac{V}{2}e^{2\alpha
\Phi}\right]\,,
\end{equation}
where  $R$ is the Ricci scalar derived from the metric $g_{\mu\nu}$
($\mu,\nu=0,1,2,3$), $g$ is the determinant of $g_{\mu\nu}$, and $\Phi$ is a
real scalar field. The constant $\alpha$ denotes the scalar self-coupling.
We used the abbreviation
$(\nabla\Phi)^2\equiv g^{\mu\nu}\partial_\mu\Phi\partial_\nu\Phi$. 
We also assumed that $V$ is a constant. It is known that the exponential potential
of this type can be found in effective field theories of string theory, higher
dimensional gravity, and higher derivative gravity (see, for example,
Ref.~\cite{KKST} and references there in).

We take an ansatz for the metric tensor as
\begin{equation}
ds^2=-N^2e^{6b(t)}dt^2+e^{2b(t)}d\mathbf{x}^2\,,
\end{equation}
where $t=x^0$, $d\mathbf{x}^2\equiv\sum_{\mu=i}^{3}(dx^{i})^2$, and
$N$ is the lapse function.
We also assume that the scalar field $\Phi$ is expressed by the function of $t$.
Then, we find
\begin{equation}
S\propto
\int dt\, N\left[-6\frac{\dot{b}^2}{N^2}+\frac{1}{2}\frac{\dot{\Phi}^2}{N^2}-
\frac{V}{2}e^{2\alpha\Phi}e^{6b}\right]\,,
\end{equation}
where the dot indicates the derivative with respect to time $t$ and
total derivatives have been dropped.

If we choose the following new variables and the defined constants
\begin{equation}
x(t)\equiv\sqrt{12}\left({b}+\frac{\alpha}{3}{\Phi}\right)\,,
\quad
y(t)\equiv{\Phi}+4{\alpha}{b}\,,
\end{equation}
and
\begin{equation}
\lambda\equiv\sqrt{\frac{3}{4}}\,,
\quad
U\equiv\left(1-\frac{4}{3}{\alpha^2}\right) V\,,
\end{equation}
the effective Lagrangian is now written by
\begin{equation}
L_N=N\left[-\frac{1}{2}\frac{\dot{x}^2}{N^2}-\frac{U}{2}e^{2\lambda
x}+\frac{1}{2}\frac{\dot{y}^2}{N^2}\right]\,.
\end{equation}

The variation in the lapse function leads to
$\frac{\partial L_N}{\partial N}=0$, that is,
\begin{equation}
-\frac{1}{2}{\dot{x}^2}+\frac{U}{2}e^{2\lambda
x}+\frac{1}{2}{\dot{y}^2}=0\,,
\label{HC}
\end{equation}
where we redefine $Nt\rightarrow t$, or equivalently, set $N=1$.
This is equivalent to the Hamiltonian constraint, which can be regarded as a
consequence of reparametrization invariance of $t$.
Note that, because of the reparametrization invariance of $t$, the overall
normalizations of the Lagrangians and the Hamiltonians which we encounter in this
paper are irrelevant for physics of cosmology. 

Now, from the Lagrangian
\begin{equation}
L_1=-\frac{1}{2}{\dot{x}^2}-\frac{U}{2}e^{2\lambda
x}+\frac{1}{2}{\dot{y}^2}\,,
\end{equation}
we can derive the equations of motion as
\begin{equation}
\ddot{x}-\lambda U e^{2\lambda x}=0\,,\quad
\ddot{y}=0\,.
\label{eom}
\end{equation}
Note that these equations have time-reversal invariance (under $t\rightarrow -t$).
Moreover, we obtain corresponding canonical momenta as
\begin{equation}
\pi_x=\frac{\partial L_1}{\partial\dot{x}}=-\dot{x}\,,\quad
\pi_y=\frac{\partial L_1}{\partial\dot{y}}=\dot{y}\,.
\end{equation}
Then, the Hamiltonian of the Liouville scalar model is found to be
\begin{equation}
H=\pi_x\dot{x}+\pi_y\dot{y}-L_1=-\frac{1}{2}{\pi_x^2}+\frac{U}{2}e^{2\lambda
x}+\frac{1}{2}{\pi_y^2}\,.
\end{equation}

One can easily confirm the Hamilton's equations
\begin{equation}
\dot{x}=\frac{\partial H}{\partial\pi_x}=-\pi_x\,,\quad
\dot{y}=\frac{\partial H}{\partial\pi_y}=\pi_y\,,\quad
\dot{\pi}_x=-\frac{\partial H}{\partial x}=-\lambda U e^{2\lambda x}\,,\quad
\dot{\pi}_y=-\frac{\partial H}{\partial y}=0\,,
\end{equation}
reproduce the classical equation of motion (\ref{eom}).

\subsection{An equivalent system}

Let us begin with the following set of the first-order differential equations:
\begin{eqnarray}
& &\dot{x}=\sqrt{U}e^{\lambda x}\cosh\lambda(y-y_0)\,,\quad
\dot{y}=-\sqrt{U}e^{\lambda x}\sinh\lambda(y-y_0)\,,\quad (U>0)
\label{eqep}\\
& &\dot{x}=\sqrt{|U|}e^{\lambda x}\sinh\lambda(y-y_0)\,,\quad
\dot{y}=-\sqrt{|U|}e^{\lambda x}\cosh\lambda(y-y_0)\,,\quad (U<0)
\label{eqem}
\end{eqnarray}
where $y_0$ is a constant.%
\footnote{Note that the invariance under the translation $y\rightarrow y+const.$
is realized both in the original and the equivalent systems. Thus, we can omit
$y_0$
, but we include the constant to make the rearrangement
of equations easy to understand.}

One can easily verify that,
if (\ref{eqep}) or (\ref{eqem}) holds, the Hamiltonian constraint (\ref{HC}) and
the equations of motion (\ref{eom}) are satisfied.%
\footnote{Of course, the set of equations transformed as $t\rightarrow -t$ also
satisfies the equations of motion of the original system.}
 The solutions of (\ref{eqep})
and (\ref{eqem}) are
\begin{equation}
x(t)=\frac{1}{2\lambda}\ln\frac{\pi_{y0}^2}{U\sinh^2\lambda
\pi_{y0}(t-t_0)}\,,\quad y(t)=\pi_{y0}(t-t_0)+y_0\,,\quad(U>0)
\label{solp}
\end{equation}
and
\begin{equation}
x(t)=\frac{1}{2\lambda}\ln\frac{\pi_{y0}^2}{|U|\cosh^2\lambda
\pi_{y0}(t-t_0)}\,,\quad y(t)=\pi_{y0}(t-t_0)+y_0\,,\quad(U<0)
\label{solm}
\end{equation}
respectively, where $\pi_{y0}$ and $t_0$ are integration constants.
If we regard $y_0$ as another `integration constant',
(\ref{solp}) and (\ref{solm}) are general solutions of Eqs.~(\ref{HC}) and
(\ref{eom}).
The degrees of freedom of the solutions are given by the number of integration
constants. The solutions of the original equations of motion have $2\times 2=4$
constants. Joining together with one constraint, we have $4-1=3$ constants
which are arbitrarily chosen. The first-order equations (\ref{eqep}) or
(\ref{eqem}) yields two integration constants but the choice of $y_0$ is arbitrary.
Thus, we again obtain $2+1=3$ degrees of freedom.
We can conclude that the necessary condition for the similar conversion of the
equations of motion is the existence of a `cyclic coordinate' in the original
system. In the Liouville scalar model, $y$ is the cyclic coordinate in the
minisuperspace, since $\frac{\partial H}{\partial y}=0$.

Note that the Lagrangian can be rewritten as
\begin{eqnarray}
L_1&=&-\frac{1}{2}\left[\dot{x}-\sqrt{U}e^{\lambda x}\cosh\lambda(y-y_0)\right]^2
+\frac{1}{2}\left[\dot{y}+\sqrt{U}e^{\lambda
x}\sinh\lambda(y-y_0)\right]^2\nonumber \\ &
&-\frac{d}{dt}\left[\lambda^{-1}\sqrt{U}e^{\lambda x}\cosh
\lambda(y-y_0)\right]\,,\quad(U>0)\\
L_1&=&-\frac{1}{2}\left[\dot{x}-\sqrt{|U|}e^{\lambda
x}\sinh\lambda(y-y_0)\right]^2 +\frac{1}{2}\left[\dot{y}+\sqrt{|U|}e^{\lambda
x}\cosh\lambda(y-y_0)\right]^2\nonumber \\ &
&-\frac{d}{dt}\left[\lambda^{-1}\sqrt{|U|}e^{\lambda x}\sinh
\lambda(y-y_0)\right]\,.\quad(U<0)
\end{eqnarray}

Now, we introduce a new equivalent system.
We find that Eqs.~(\ref{eqep}) and (\ref{eqem}) can also be derived
from the following first-order Lagrangian:
\begin{eqnarray}
\bar{L}&=&(y-y_0)\dot{x}-\lambda^{-1}\sqrt{U}e^{\lambda x}\sinh\lambda(y-y_0)\,,
\quad(U>0)\\
\bar{L}&=&(y-y_0)\dot{x}-\lambda^{-1}\sqrt{|U|}e^{\lambda x}\cosh\lambda(y-y_0)\,.
\quad(U<0)
\end{eqnarray}
According to the prescription of Faddeev and Jackiw \cite{FJ,Jackiw},
the Hamiltonian of this equivalent system is
\begin{eqnarray}
\bar{H}&=&\lambda^{-1}\sqrt{U}e^{\lambda x}\sinh\lambda p\,,\quad(U>0)
\label{lhp}\\
\bar{H}&=&\lambda^{-1}\sqrt{|U|}e^{\lambda x}\cosh\lambda p\,,\quad(U<0)
\label{lhm}
\end{eqnarray}
where $p$ is the conjugate momentum of $x$.

The Hamilton's equations are found to be
\begin{eqnarray}
& &\dot{x}=\frac{\partial\bar{H}}{\partial p}=\sqrt{U}e^{\lambda x}\cosh\lambda
p\,,\quad
\dot{p}=-\frac{\partial\bar{H}}{\partial x}=-\sqrt{U}e^{\lambda x}\sinh\lambda
p\,,\quad(U>0)\\
& &\dot{x}=\frac{\partial\bar{H}}{\partial p}=\sqrt{|U|}e^{\lambda x}\sinh\lambda
p\,,\quad
\dot{p}=-\frac{\partial\bar{H}}{\partial x}=-\sqrt{|U|}e^{\lambda x}\cosh\lambda
p\,,\quad(U<0)
\end{eqnarray}
which are surely equivalent to (\ref{eqep}) and (\ref{eqem}), respectively,
provided that we regard
$p\approx y-y_0$ (where ``$\approx$'' means ``is identified with'').

The idea of our prescription originates from the canonical transformation of
the Liouville Hamiltonian \cite{Ghandour}.
If the Liouville Hamiltonian reduces to the form of $\frac{1}{2}\Pi^2$ (see
Appendix \ref{AA}), we can choose the Hamiltonian constraint as
$\Pi\pm\pi_y=0$, instead of
$-\Pi^2+\pi_y^2=0$. Our further step is to propose that the classical constraint is
expressed in the first-order Lagrangian \`a la Faddeev and Jackiw \cite{FJ,Jackiw}
and then we can obtain the equivalent Hamiltonian $\bar{H}$ of a reduced set of
canonical variables, $x$ and
$p$. Thus, $y-y_0\approx p$ plays a role of a universal `clock'.

In the next section, we show another model, the conformal scalar model
and its equivalent Hamiltonian.
The study of quantum dynamics of two models will be described in Sec.~\ref{sec4}
and Sec.~\ref{sec5}.

\section{the conformal scalar model}
\label{sec3}

\subsection{The original system}

The simplest model for a spatially homogeneous and isotropic
Universe is described by the action \cite{HH,Hawking,Kiefer2}
\begin{equation}
S=\int d^4x\sqrt{-g}\left[R-\frac{1}{2}(\nabla\phi)^2-\frac{1}{12}R\phi^2\right]\,,
\end{equation}
where $\phi$ is a
real scalar field conformally coupled to the scalar curvature. 
We consider this system in the minisuperspace as the second example of the
integrable model in this paper and call it the conformal scalar model.

We take an ansatz for the metric tensor as
\begin{equation}
ds^2=a^2(t)\left[-N^2
dt^2+\frac{d\rho^2}{1-k\rho^2}+\rho^2(d\theta^2+\sin^2\theta d\psi^2)\right]\,,
\end{equation}
where $k (=\pm 1)$
 indicates the sign of the spatial curvature.%
\footnote{For the case of $k=0$ is rather trivial, we do not deal with the case,
in this paper.} We also assume that the conformally invariant scalar field $\phi$
is expressed by the function of
$t$. Then, we find
\begin{equation}
S\propto
12\int dt\,
N\left[-\frac{1}{2}\frac{\dot{a}^2}{N^2}+\frac{k}{2}a^2+\frac{1}{2}\frac{\dot{\chi}^2}{N^2}-
\frac{k}{2}\chi^2\right]\,,
\end{equation}
where $\chi=a\phi/\sqrt{12}$ and total derivatives have been dropped.
By using new variables
\begin{equation}
a(t)=r(t)\cosh\varphi(t)\,,\quad\chi(t)=r(t)\sinh\varphi(t)\,,
\end{equation}
we find a simple form of the Lagrangian
\begin{equation}
L_1=-\frac{1}{2}\dot{r}^2+\frac{k}{2}r^2+\frac{1}{2}r^2\dot{\varphi}^2\,,
\label{l2}
\end{equation}
and the constraint
\begin{equation}
-\frac{1}{2}\dot{r}^2-\frac{k}{2}r^2+\frac{1}{2}r^2\dot{\varphi}^2=0\,.
\label{cl}
\end{equation}
Note that the variable $\varphi$ is a `cyclic coordinate' in the two dimensional
minisuperspace.

The equations of motion derived from the Lagrangian (\ref{l2}) are
\begin{equation}
\ddot{r}+r\dot{\varphi}^2+kr=0\,,\quad
\ddot{\varphi}+2\frac{\dot{r}}{r}\dot{\varphi}=0\,.
\label{eom2}
\end{equation}
Note that these equations have time-reversal invariance.

\subsection{An equivalent system and canonical transformations}

Now, we find that the set of equations
\begin{eqnarray}
& &\dot{r}=r\sinh 2(\varphi-\varphi_0)\,,\quad\dot{\varphi}=-\cosh
2(\varphi-\varphi_0)\,,
\quad \mbox{for } k=1\,,
\label{kp}
\\
& &\dot{r}=r\cosh 2(\varphi-\varphi_0)\,,\quad\dot{\varphi}=-\sinh
2(\varphi-\varphi_0)\,,
\quad \mbox{for } k=-1\,,
\label{km}
\end{eqnarray}
reproduces the equations of motion (\ref{eom2}) and the constraint (\ref{cl}).
Here, $\varphi_0$ is a constant.
Note that the original Lagrangian can be written as
\begin{eqnarray}
L_1&=&-\frac{1}{2}\left[\dot{r}-r\sinh 2(\varphi-\varphi_0)\right]^2
+\frac{1}{2}r^2\left[\dot{\varphi}+\cosh 2(\varphi-\varphi_0)\right]^2
\nonumber \\
& &-\frac{d}{dt}\left[\frac{r^2}{2}\sinh 2(\varphi-\varphi_0)\right]\,,
\quad\mbox{for } k=1\,,\\
L_1&=&-\frac{1}{2}\left[\dot{r}-r\cosh 2(\varphi-\varphi_0)\right]^2
+\frac{1}{2}r^2\left[\dot{\varphi}+\sinh 2(\varphi-\varphi_0)\right]^2
\nonumber \\
& &-\frac{d}{dt}\left[\frac{r^2}{2}\cosh 2(\varphi-\varphi_0)\right]\,,
\quad\mbox{for } k=-1\,.
\end{eqnarray}

The Lagrangians which lead to the first-order equations (\ref{kp}) and (\ref{km})
are
\begin{eqnarray}
\bar{L}&=&(\varphi-\varphi_0) r\dot{r}-\frac{r^2}{2}\cosh 2(\varphi-\varphi_0)
=(\varphi-\varphi_0)\dot{q}-q\cosh 2(\varphi-\varphi_0)\,,~~\mbox{for }
k=1\,,\\
\bar{L}&=&(\varphi-\varphi_0) r\dot{r}-\frac{r^2}{2}\sinh 2(\varphi-\varphi_0)
=(\varphi-\varphi_0)\dot{q}-q\sinh 2(\varphi-\varphi_0)\,,~~\mbox{for }
k=-1\,,
\end{eqnarray}
respectively, where we set $q\equiv r^2/2$.
Now, we can obtain the classical Hamiltonians with the aid of the Faddeev--Jackiw
method as in the previous section:
\begin{eqnarray}
\bar{H}&=&q\cosh 2p\,,\quad\mbox{for } k=1\,,\\
\bar{H}&=&q\sinh 2p\,,\quad\mbox{for } k=-1\,.
\end{eqnarray}

In this model, this is not the final form of physical interest.
One can find the Poisson's bracket $\{Q, P\}=1$, where
\begin{equation}
Q\equiv\sqrt{2q}\cosh p\,,\quad P\equiv\sqrt{2q}\sinh p\,.
\end{equation}
Using the new canonical set of variables, the equivalent Hamiltonians read
\begin{eqnarray}
\bar{H}&=&\frac{1}{2}\left(P^2+Q^2\right)\,,\quad\mbox{for } k=1\,,\\
\bar{H}&=&QP\,,\quad\mbox{for } k=-1\,.
\end{eqnarray}
To check the validity, we derive the Hamilton's equations
\begin{eqnarray}
& &\dot{Q}=\frac{\partial\bar{H}}{\partial P}= P\,,\quad
\dot{P}=-\frac{\partial\bar{H}}{\partial Q}=-Q\,,
\quad\mbox{for } k=1\,,\\
& &\dot{Q}=\frac{\partial\bar{H}}{\partial P}=Q\,,\quad
\dot{P}=-\frac{\partial\bar{H}}{\partial Q}=-P\,,
\quad\mbox{for } k=-1\,.
\end{eqnarray}
Note that $p$ was identified with $\varphi-\varphi_0$
by the Faddeev--Jackiw method. Hereafter, we again denote the identification by
using the symbol ``$\approx$'' for conciseness.  Then, we find
\begin{eqnarray}
& &Q\approx
r\cosh(\varphi-\varphi_0)=r(\cosh\varphi\cosh\varphi_0-
\sinh\varphi\sinh\varphi_0)=a\cosh\varphi_0-\chi\sinh\varphi_0\,,\\
& &P\approx
r\sinh(\varphi-\varphi_0)=r(\sinh\varphi\cosh\varphi_0-
\cosh\varphi\sinh\varphi_0)=\chi\cosh\varphi_0-a\sinh\varphi_0\,,
\label{relations}
\end{eqnarray}
or
\begin{equation}
a\approx
Q\cosh\varphi_0+P\sinh\varphi_0\,,
\quad\chi\approx
P\cosh\varphi_0+Q\sinh\varphi_0\,.
\label{relations1}
\end{equation}
After straightforward calculations, one can find that the Hamilton's equations are
equivalent to the first-order equations (\ref{kp}) and (\ref{km}), as far as we
regard
$p\approx \varphi-\varphi_0$.

The classical solutions of the systems are found to be
\begin{equation}
Q(t)=Q_0\cos t+P_0\sin t\,,\quad
P(t)=P_0\cos t-Q_0\sin t\,,
\quad\mbox{for } k=1\,,
\label{cqp}
\end{equation}
\begin{equation}
Q(t)=Q_0\,e^t\,,\quad
P(t)=P_0\,e^{-t}\,,
\quad\mbox{for } k=-1\,,
\label{cqm}
\end{equation}
where $Q_0$ and $P_0$ are constants.

It is remarkable that the conformal scalar model with $k=1$ reduces to
a system of a harmonic oscillator while the model with $k=-1$ reduces to the
well-known $xp$ model \cite{BK99a,BK99b,Conne,Sierra,ST,BBM}, whose spectrum is
considered to be related to the zeros of the Riemann zeta function.
Incidentally, it is also known that the similar Hamiltonian has been studied in the
model of loop quantum cosmology \cite{Bojowald1,Bojowald2}.

For the conformal scalar model with $k=-1$, a remarkable canonical transformation
remains to be done. The new canonical pair is
\begin{equation}
Q'\equiv\frac{Q-P}{\sqrt{2}}\,,\quad P'\equiv\frac{P+Q}{\sqrt{2}}\,.
\end{equation}
Then, the Hamiltonian is rewritten as
\begin{equation}
\bar{H}=\frac{1}{2}({P'}^2-{Q'}^2)\,.
\label{asham}
\end{equation}
This is just a so-called inverted harmonic oscillator.%
\footnote{Therefore, the inverted harmonic oscillator may also
have some concern with the Riemann zeta function \cite{BKL,BKRT,Khare}.} 
The classical
solution of the system can be written as
\begin{equation}
Q'(t)=Q'_0\cosh t+P'_0\sinh t\,,\quad
P'(t)=P'_0\cosh t+Q'_0\sinh t\,,
\end{equation}
where $Q'_0$ and $P'_0$ are constants.
Note that the correspondence to the original variables is
\begin{equation}
Q'\approx\frac{1}{\sqrt{2}}(a-\chi)e^{\varphi_0}\,,
\quad P'\approx\frac{1}{\sqrt{2}}(a+\chi)e^{-\varphi_0}\,.
\label{relations2}
\end{equation}

In the next two sections, we will try to investigate the quantum nature of our
models.

\section{towards quantum dynamics of the Liouville scalar model}
\label{sec4}

If we introduce 
the wave function of the Universe $\Psi(t, q)$, the Schr\"odinger equation
can be written as
\begin{equation}
i\hbar\frac{\partial\Psi}{\partial t}=\hat{H}\Psi\,,
\label{seq}
\end{equation}
where $\hbar$ is the Planck's constant.%
\footnote{In this section, we illustrate the Planck's constant, because we want to
emphasize the aspect of quantum effects.} 
Here $\hat{H}$ is the `quantum Hamiltonian', usually connected to the classical
Hamiltonian $H(q, p)\rightarrow \hat{H}(\hat{q},\hat{p})$, up to the operator
ordering.

As a wave equation obtained by replacing
$\hat{p}$ with $\frac{\hbar}{i}\frac{\partial}{\partial q}$,%
\footnote{It is notable that the noncommutativity of two minisuperspace variables 
naturally emerges,
since $y-y_0\approx p$, which is now the conjugate variable of $x$, in the
Liouville scalar model. For noncommutative quantum cosmology, 
in which original dynamical variables do not commute,
see
\cite{COR,KKST2}, for example.} 
the Schr\"odinger equation for the Liouville scalar
model described in Sec.~\ref{sec2}  is hard to solve, because there appear
infinite derivatives. We postpone the study on the wave function itself for the
future work
, here we consider the quantum effects on the
classical solutions in the Liouville scalar model.
The conformal scalar model will be studied in the next section.

It is well-known that the Ehrenfest's theorem tells
\begin{equation}
\frac{d\langle \hat{A}\rangle}{dt}=\frac{1}{i\hbar}\langle[\hat{A},
\hat{H}]\rangle\,,
\end{equation}
where $\langle \hat{A}\rangle$ is the expectation value of the operator $\hat{A}$.
The approximation of the equation in the order $\hbar^0$ gives the classical
equations of motion. To evaluate the quantum corrections, we should treat the
fluctuation operator
$\delta\hat{A}=\hat{A}-\langle\hat{A}\rangle$ appropriately.

Recently, Shigeta et al.\cite{SMH,Shigeta} defined an expression for the
expectation value by means of cumulants among the canonical pair of the variables
$q$ and $p$. That is defined as
\begin{equation}
\langle A_S(\hat{q}, \hat{p})\rangle_M\equiv
\exp\left(\sum_{m=2}^M\sum_{0\le\ell\le
m}\frac{\kappa_{\ell,m-\ell}}{\ell!(m-\ell)!}
\frac{\partial_m}{\partial q^\ell\partial p^{m-\ell}}\right)A(q,
p)
\,,
\end{equation}
where we introduced the cumulants $\kappa_{i,j}\equiv
\langle((\delta\hat{q})^i(\delta\hat{p})^j)_S\rangle$ $(i,j=0,1,2,\ldots)$ and the
subscript
$S$ indicates a symmetric sum of the operators defined as
$\langle((\delta\hat{q})^i(\delta\hat{p})^j)_S\rangle\equiv\frac{1}{2}
\langle((\delta\hat{q})^i(\delta\hat{p})^j+
(\delta\hat{p})^j(\delta\hat{q})^i)\rangle$.
This definition can be expressed as
$\frac{1}{2}\{(\delta\hat{q})^i, (\delta\hat{p})^j\}$,
where $\{~,~\}$ denotes the anticommutator.

Here, we takes the approximation by taking $M=2$ and we set the approximated
Hamiltonian as
\begin{equation}
\tilde{H}(q, p; \kappa_{2,0}, \kappa_{1,1},
\kappa_{0,2})\equiv\langle\hat{H}(\hat{q}, \hat{p})\rangle_2\,.
\end{equation}
Then, the following simultaneous equations are derived from the Ehrenfest's
theorem:\cite{SMH,Shigeta}
\begin{eqnarray}
\dot{q}&=&\tilde{H}^{(0,1)}\,,
\label{simeq1}\\
\dot{p}&=&-\tilde{H}^{(1,0)}\,,
\label{simeq2}\\
\dot{\kappa}_{2,0}&=&2\kappa_{2,0}\tilde{H}^{(1,1)}+2\kappa_{1,1}\tilde{H}^{(0,2)}\,,
\label{simeq3}\\
\dot{\kappa}_{1,1}&=&-\kappa_{2,0}\tilde{H}^{(2,0)}+\kappa_{0,2}\tilde{H}^{(0,2)}\,,
\label{simeq4}\\
\dot{\kappa}_{0,2}&=&-2\kappa_{1,1}\tilde{H}^{(2,0)}-2\kappa_{0,2}\tilde{H}^{(1,1)}\,,
\label{simeq5}
\end{eqnarray}
where $\tilde{H}^{(2,0)}\equiv\frac{\partial^2\tilde{H}}{\partial q^2}$,
$\tilde{H}^{(1,1)}\equiv\frac{\partial^2\tilde{H}}{\partial q\partial p}$,
and
$\tilde{H}^{(0,2)}\equiv\frac{\partial^2\tilde{H}}{\partial p^2}$.

Accordingly, the approximated Hamiltonian for the Liouville scalar model is
calculated as
\begin{eqnarray}
\tilde{H}(x,p;\kappa_{2,0},\kappa_{1,1},\kappa_{0,2})&=&
\frac{\sqrt{U}}{\lambda}\exp\left[\lambda
x+\frac{\lambda^2(\kappa_{2,0}+\kappa_{0,2})}{2}\right]
\sinh(\lambda p+\lambda^2\kappa_{1,1})\,,(U>0)
\label{Rsim1}
\\
\tilde{H}(x,p;\kappa_{2,0},\kappa_{1,1},\kappa_{0,2})&=&
\frac{\sqrt{|U|}}{\lambda}\exp\left[\lambda
x+\frac{\lambda^2(\kappa_{2,0}+\kappa_{0,2})}{2}\right]
\cosh(\lambda p+\lambda^2\kappa_{1,1})\,,(U<0)
\label{Rsim2}
\end{eqnarray}
where the canonical pair is replaced by $(q,p)\rightarrow(x,p)$.

Now, we can numerically solve the simultaneous equations
(\ref{simeq1}-\ref{simeq5}) (with $q\rightarrow x$). Since we know that $p\approx
y-y_0$ evolves linearly in time in the classical system, we illustrate evolutions
of variables as functions of $p$. 

\begin{figure}[ht]
\centering
\includegraphics[width=7cm]{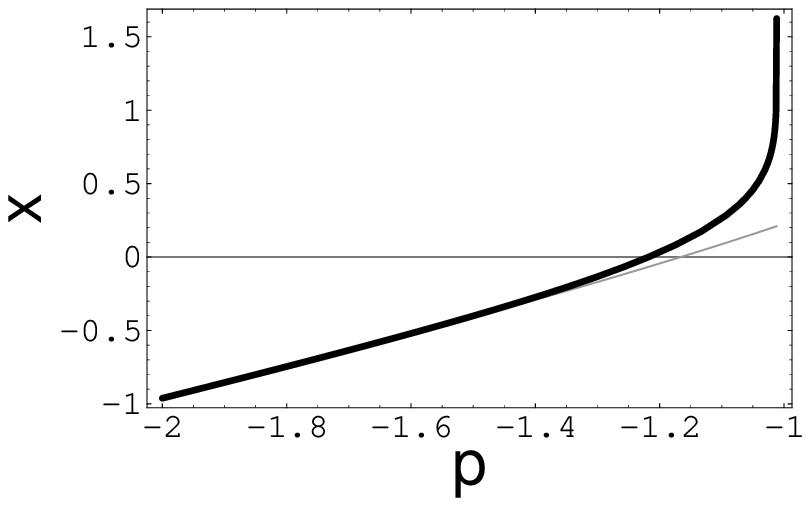}\quad\quad
\includegraphics[width=7.5cm]{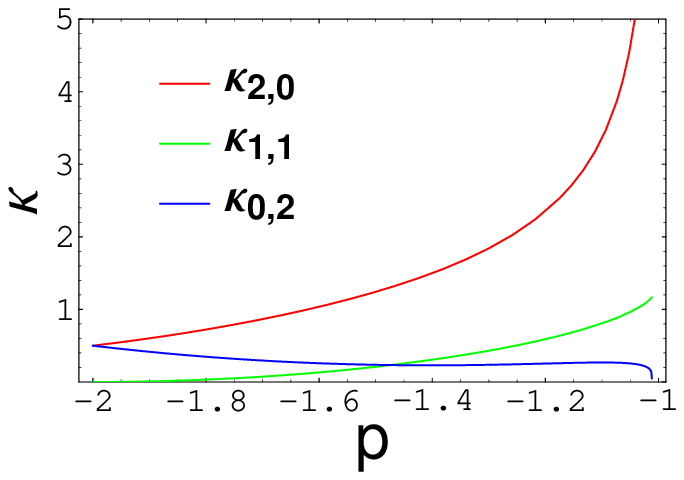}
\\
(a) \hspace{7.5cm} (b)
\caption{(a) The quantum corrected solution (black curve) for
(\ref{simeq1}-\ref{simeq5}) and the classical solution (gray curve) under the same
initial conditions are plotted in the Liouville scalar model with $U>0$. For the
choice of initial values and constants, see the text. (b) The values of the
cumulants are plotted against
$p$ in the Liouville scalar model with $U>0$. For the choice of initial values and
constants, see the text.}
\label{fig1}
\end{figure}

Fig.~\ref{fig1} shows the classical solutions and quantum corrections in the
Liouville scalar model with $U>0$.
We set $\hbar=1$, $\lambda=\sqrt{3}/2$, and $U=3/4$.
Initial conditions are given at $p=p_0=-2$. We choose $x(p_0)=-0.96163$,
$\kappa_{2,0}(p_0)=\kappa_{0,2}(p_0)=1/2$, and $\kappa_{1,1}(p_0)=0$.

Note that $\kappa_{2,0}\kappa_{0,2}=1/4$ is the minimal value which comes from the
uncertainty principle  (i.e., we assume a coherent state initially).
Note also that $\kappa_{2,0}\kappa_{0,2}-\kappa_{1,1}^2$ is a constant
\cite{SMH,Shigeta}.

In Fig.~\ref{fig1} (a), the quantum corrected solution
(black curve) for (\ref{simeq1}-\ref{simeq5}) and the classical solution (gray
curve) under the same initial condition are plotted. Quantum corrections enhances
the value of $x$. In Fig.~\ref{fig1} (b), the value of the cumulants are plotted.
Around $p\fallingdotseq -1$, the value of $\kappa_{0,2}$ tends to take a minus
value and $\kappa_{2,0}$ grows unlimitedly.
This indicates that the point is the limit of the present approximation scheme
and we should incorporate higher-order cumulants to obtain precise quantum effects.

\begin{figure}[ht]
\centering
\includegraphics[width=7cm]{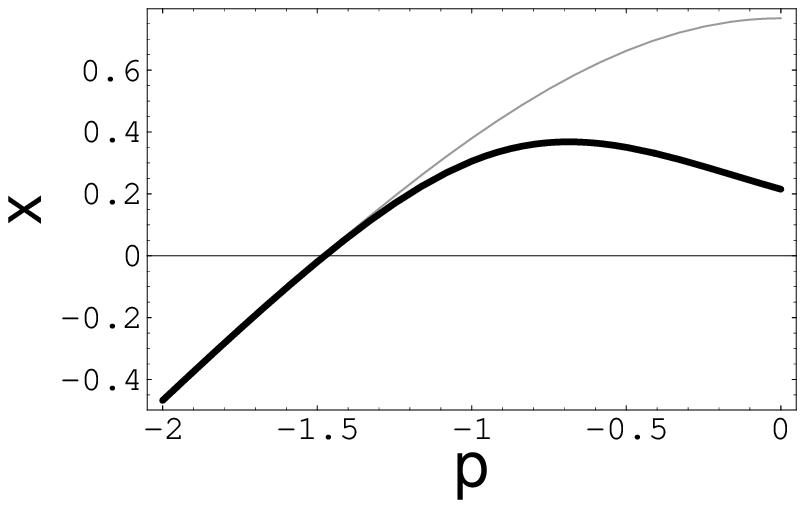}\quad\quad
\includegraphics[width=7.5cm]{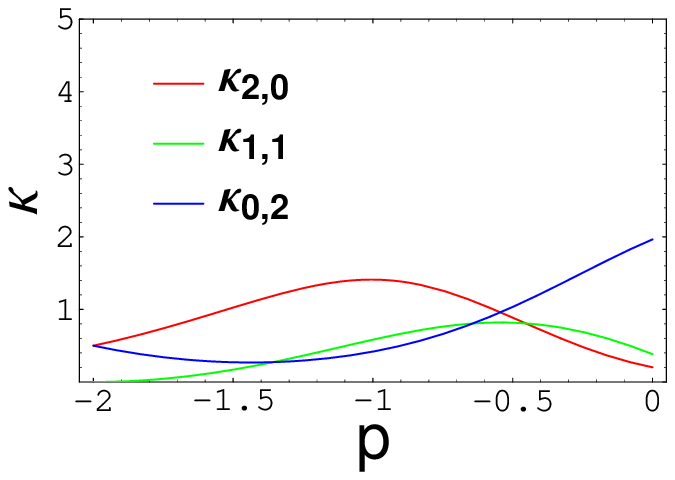}
\\
(a) \hspace{7.5cm} (b)
\caption{(a) The quantum corrected solution (black curve) for
(\ref{simeq1}-\ref{simeq5}) and the classical solution (gray curve) under the same
initial conditions are plotted in the Liouville scalar model with $U<0$. For the
choice of initial values and constants, see the text. (b) The value of the
cumulants are plotted against
$p$ in the Liouville scalar model with $U<0$. For the choice of initial values and
constants, see the text.}
\label{fig2}
\end{figure}

Fig.~\ref{fig2} shows the classical solutions and quantum corrections in the
Liouville scalar model with $U<0$.
We set $\hbar=1$, $\lambda=\sqrt{3}/2$, and $U=-1/2$.
Initial conditions are given at $p=p_0=-2$. We choose $x(p_0)=-0.467658$,
$\kappa_{2,0}(p_0)=\kappa_{0,2}(p_0)=1/2$, and $\kappa_{1,1}(p_0)=0$.

In Fig.~\ref{fig2} (a), the quantum corrected solution
(black curve) for (\ref{simeq1}-\ref{simeq5}) and the classical solution (gray
curve) under the same initial conditions are plotted in the model with $U<0$.
Quantum corrections reduces the value of $x$. In Fig.~\ref{fig2} (b), the value
of the cumulants are plotted. The evolution of the cumulants in the model with
$U<0$ are more moderate than the model with $U>0$.

In both cases, we should consider various cases of squeezed states, i.e.,
$\kappa_{2,0}\gg\kappa_{0,2}\mbox{ or }\kappa_{2,0}\ll \kappa_{0,2}$, and study the
evolution of decoherence in the higher-order approximation, in order to
conclude the precise behavior of the quantum Liouville cosmology.


Here, we would like to comment on the realization of the original Hamiltonian
constraint (\ref{HC}) in the current equivalent system.
From Eqs.~(\ref{simeq1}-\ref{Rsim2}), we find that $\dot{x}^2-\dot{y}^2=
U\exp[2\lambda x+\lambda^2(\kappa_{2,0}+\kappa_{0,2})]$.
Thus, the constraint is satisfied semiclassically in the sense of the
Ehrenfest's theorem, but the quantum fluctuations (cumulants) may cause deviations.
Further study on the constraint in the system is very important and we wish to
report on the study in the future.


Before closing this section, we place a comment on a further canonical
transformation. Incidentally, if we perform canonical transformation
\begin{equation}
X\equiv\frac{x-p}{\sqrt{2}}\,,\quad P\equiv\frac{x+p}{\sqrt{2}}\,,
\end{equation}
on the equivalent Hamiltonians (\ref{lhp}) and (\ref{lhm}),
we obtain 
\begin{eqnarray}
\bar{H}&=&(2\lambda)^{-1}\sqrt{U}[e^{\sqrt{2}\lambda P}-e^{\sqrt{2}\lambda
X}]\,,\quad(U>0)
\\
\bar{H}&=&(2\lambda)^{-1}\sqrt{|U|}[e^{\sqrt{2}\lambda P}+e^{\sqrt{2}\lambda
X}]\,,\quad(U<0)
\end{eqnarray}
as the Hamiltonians. Similar types of Hamiltonians
have been studied in an area of mathematical physics recently, see for example
\cite{FT,GHM,LST,CMS,MZ,GM}.
We should further study the Liouville scalar model of quantum universe
in the future work,
considering the help of various possible ways to investigate it.

\section{quantum dynamics and wave function of the conformal scalar model}
\label{sec5}

\subsection{Cumulant quantum dynamics of the conformal scalar model}

Incidentally, for the conformal scalar model, we find the quantum Hamiltonian
formulated in the previous section as
\begin{equation}
\tilde{H}(Q,P;\kappa_{2,0},\kappa_{1,1},\kappa_{0,2})=
\frac{1}{2}\left(P^2+\kappa_{0,2}+Q^2+\kappa_{2,0}\right)\,,\quad(k=1)
\end{equation}
\begin{equation}
\tilde{H}(Q,P;\kappa_{2,0},\kappa_{1,1},\kappa_{0,2})=
\left(QP\right)_S+\kappa_{1,1}\,.\quad(k=-1)
\end{equation}
Note that these are also exact quantum Hamiltonians up to the all order $M=\infty$.
It can be seen from the quantum Hamiltonians and the simultaneous equations
(\ref{simeq1}-\ref{simeq5}) in which the dynamics of the expectation values of $Q$
and
$P$ and the dynamics of the cumulants are completely decoupled in the conformal
scalar model in the both cases $k=1$ and $k=-1$. Thus, the solutions for $Q$ and
$P$ coincide with the classical solutions (\ref{cqp}) and (\ref{cqm}) for $k=1$ and
$k=-1$, respectively. The solutions for the cumulants are
$\kappa_{2,0}=A-B\cos 2(t-t_1)$, $\kappa_{0,2}=A+B\cos 2(t-t_1)$,
and $\kappa_{1,1}=B\sin 2(t-t_1)$ for $k=1$,%
\footnote{In the case of a coherent state, the constant $B$ equals to zero.} 
while $\kappa_{2,0}=A\exp[2(t-t_1)]$,
$\kappa_{0,2}=A\exp[-2(t-t_1)]$, and $\kappa_{1,1}=B$ for $k=-1$.
We used $A$, $B$, and $t_1$ as integration constants here.
Therefore, we find that the magnitude of coherence of the initial state becomes
oscillatory in the conformal scalar model with $k=1$ while the decoherence
develops in the conformal scalar model with $k=-1$.

\subsection{The wave function and the Wigner function of the conformal scalar model
with
$k=1$}

Especially, since an arbitrary state of the conformal scalar model with $k=1$
can be expressed by the states of a harmonic oscillator, 
a wave packet corresponding to a semiclassical evolution can be described by
superposition of eigenstates. Particularly, a coherent state is interesting as an
initial state.

We can analyze the conformal scalar model with $k=1$, which is the simplest toy
model of quantum cosmology,  by using the known results on harmonic oscillators.
According to a typical wave packet exhibited in Ref.~\cite{RB}, a typical wave
function in the model with
$k=1$ can be written as
\begin{eqnarray}
\Psi(Q,t)&=&\frac{1}{\sqrt{\pi^{1/2}A(t)}}\exp\left[\frac{i \{(Q^2+Q_0^2)\cos
t-2 Q_0Q\}}{2\hbar
\sin t}\right]\nonumber \\
& &\times
\exp\left[-\frac{i\beta}{2\hbar\sin t}\frac{(Q-Q_0\cos t-P_0\sin
t)^2}{A(t)}\right]\,,
\end{eqnarray}
where
\begin{equation}
A(t)\equiv\beta\cos t+i\frac{\hbar}{\beta}\sin t\,,
\end{equation}
and $Q_0$, $P_0$, and $\beta$ are constants.
The probability density is given by
\begin{equation}
|\Psi(Q,t)|^2=\frac{1}{\sqrt{\pi}|A(t)|}
\exp\left[-\frac{(Q-Q_0\cos t-P_0\sin t)^2}{|A(t)|^2}\right]\,.
\end{equation}
The initial wave packet gives $\langle Q\rangle=Q_0$ for $t=0$.
Note that the second order cumulant $\kappa_{2,0}$ is given by
$|A(t)|^2=\frac{\hbar}{2}\left(\frac{\beta^2}{\hbar}\cos^2t+
\frac{\hbar}{\beta^2}\sin^2t\right)$.
The initial condition on the coherence is given by the constant $\beta$
and $\beta=\sqrt{\hbar}$ gives a coherent state and then $|A(t)|^2=\hbar/2$.


Generally speaking, the Wigner function \cite{Wigner,Case,WF} is defined, in terms
of a wave function
$\phi(q)$, by
\begin{equation}
W(q, p)\equiv \frac{1}{2\pi\hbar}\int_{-\infty}^\infty du\,
\phi^*\left(q-\frac{u}{2}\right)\phi\left(q+\frac{u}{2}\right)e^{-i\frac{p}{\hbar}u}\,.
\end{equation}
The Wigner function has beautiful properties, such as
\begin{equation}
\int_{-\infty}^\infty dp\,W(q,
p)=\left|\phi\left(q\right)\right|^2\,,\quad
\int_{-\infty}^\infty dq\,W(q,
p)=|\tilde{\phi}\left(p\right)|^2\,,
\end{equation}
where $\tilde{\phi}(p)$ is the Fourier transform of $\phi(q)$.
Note that the Wigner function itself is not positive definite in general.


For the Gaussian wave packet of a harmonic oscillator,
the simple result of the Wigner function has already been known \cite{Case}.
In the present model, the Wigner function is given by
\begin{equation}
W(Q, P,t)=\frac{1}{\pi\hbar}
\exp\left[-\frac{1}{\beta^2}(Q\cos t-P\sin
t-Q_0)^2-\frac{\beta^2}{\hbar^2}(P\cos t+Q\sin
t-P_0)^2\right]\,.
\end{equation}
It is remarkable that the Wigner function is positive in this case.

Remembering the relation between $(Q, P)$ and $(a, \chi)$ (\ref{relations1})
in Sec.~\ref{sec3}, we can plot the Wigner function as a two-parameter function of
$a$ and $\chi$. We set $\hbar=1$, and $
a
(0)=0$ and $
\chi
(0)=5$ in the following figures. Fig.~\ref{fig3} shows the Wigner
functions of (a) a coherent state,
$\beta=1$, (b) a squeezed state, $\beta=2$ and (c) a squeezed state, $\beta=1/2$,
for $\varphi_0=0$. 
Fig.~\ref{fig4} shows the Wigner functions of the same states
for $\varphi_0=0.3$. 
Fig.~\ref{fig5} shows the Wigner functions of the same states
for $\varphi_0=-0.3$. 
In each figure, the Wigner function at times $t=0$, $t=\pi/2$, $t=\pi$, and
$t=3\pi/2$ are shown at once.

\begin{figure}[ht]
\centering
\includegraphics[width=5cm]{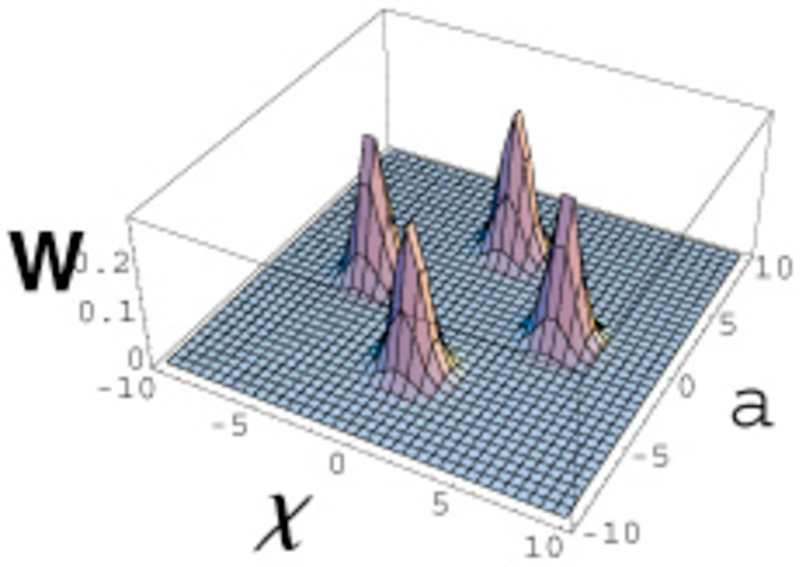}\quad
\includegraphics[width=5cm]{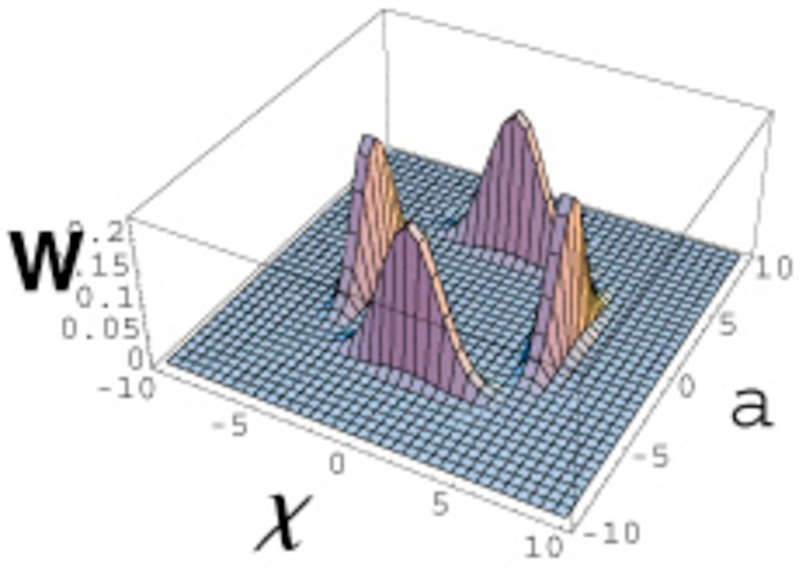}\quad
\includegraphics[width=5cm]{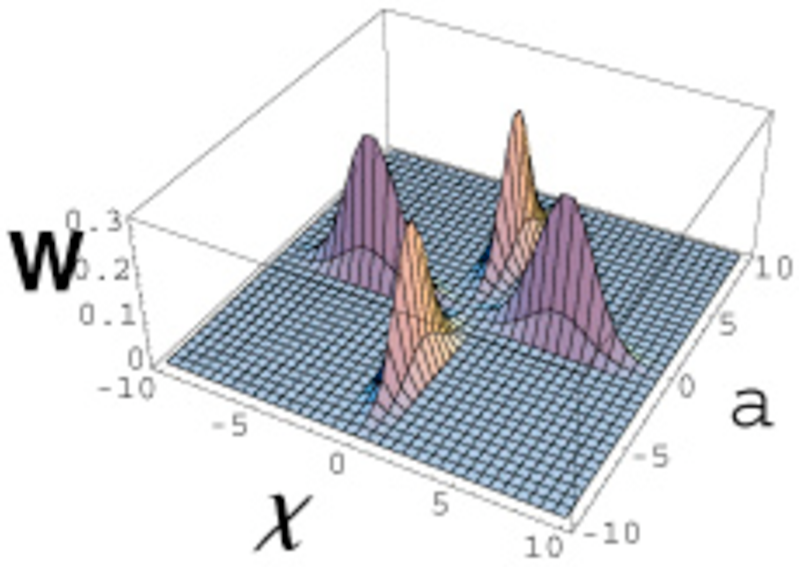}
\\
(a) \hspace{5cm} (b) \hspace{5cm} (c)
\caption{Plots of the Wigner functions of (a) a coherent state, $\beta=1$,
(b) a squeezed state, $\beta=2$ and (c) a squeezed state, $\beta=1/2$,
for $\varphi_0=0$ in the conformal scalar model with $k=1$. In each figure, the
Wigner function at times
$t=0$,
$t=\pi/2$,
$t=\pi$, and
$t=3\pi/2$ are shown at once.}
\label{fig3}
\end{figure}
\begin{figure}[ht]
\centering
\includegraphics[width=5cm]{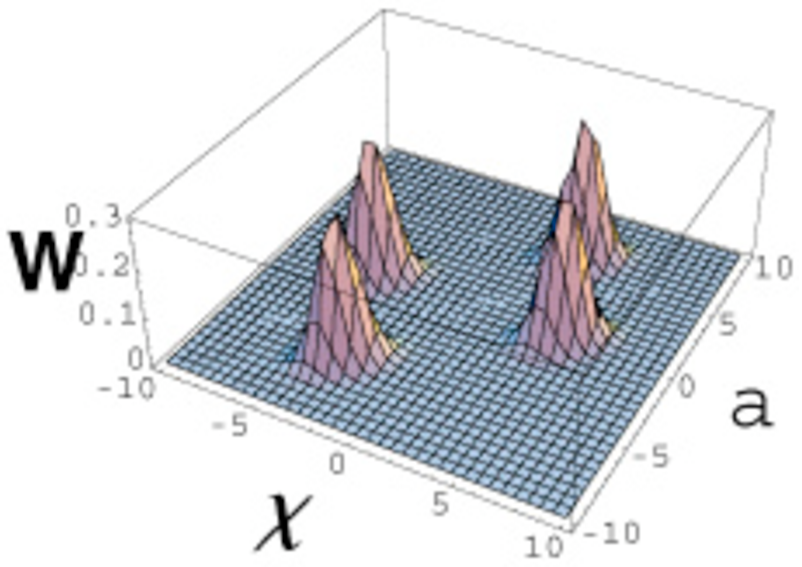}\quad
\includegraphics[width=5cm]{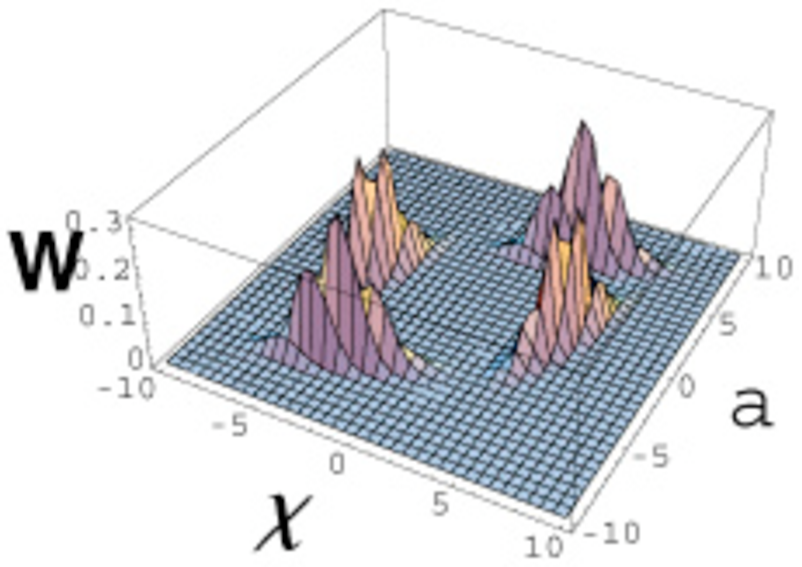}\quad
\includegraphics[width=5cm]{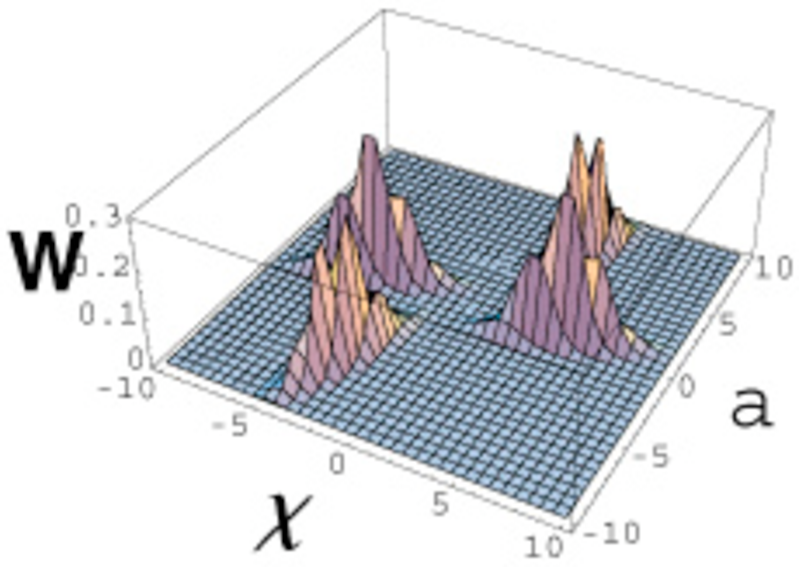}
\\
(a) \hspace{5cm} (b) \hspace{5cm} (c)
\caption{Plots of the Wigner functions of (a) a coherent state, $\beta=1$,
(b) a squeezed state, $\beta=2$ and (c) a squeezed state, $\beta=1/2$,
for $\varphi_0=0.3$ in the conformal scalar model with $k=1$. In each figure, the
Wigner function at times $t=0$, $t=\pi/2$, $t=\pi$, and
$t=3\pi/2$ are shown at once.}
\label{fig4}
\end{figure}
\begin{figure}[ht]
\centering
\includegraphics[width=5cm]{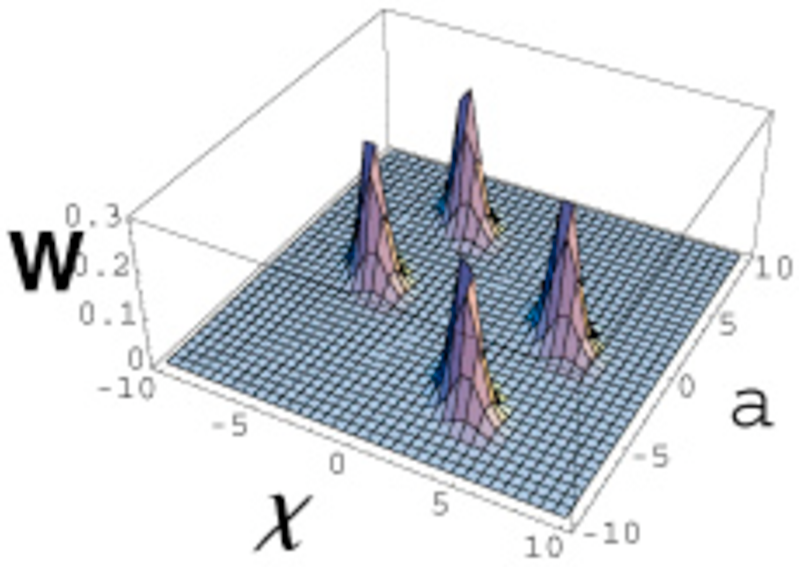}\quad
\includegraphics[width=5cm]{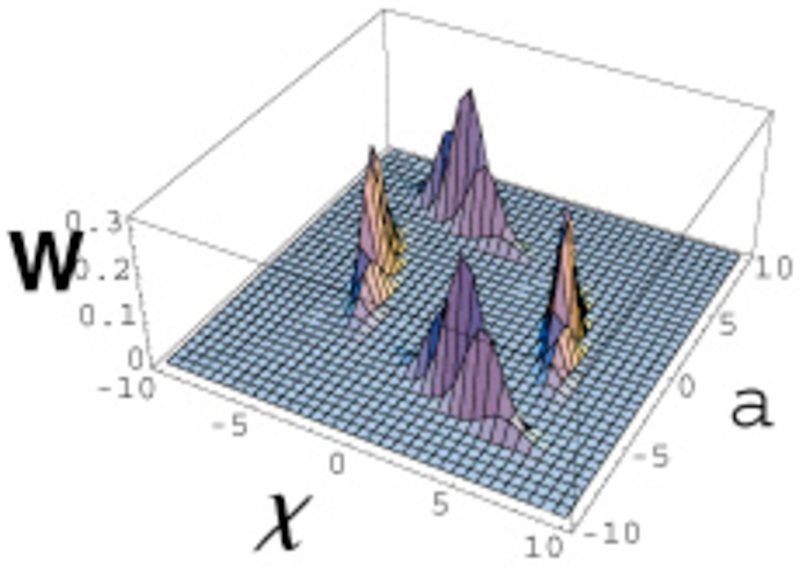}\quad
\includegraphics[width=5cm]{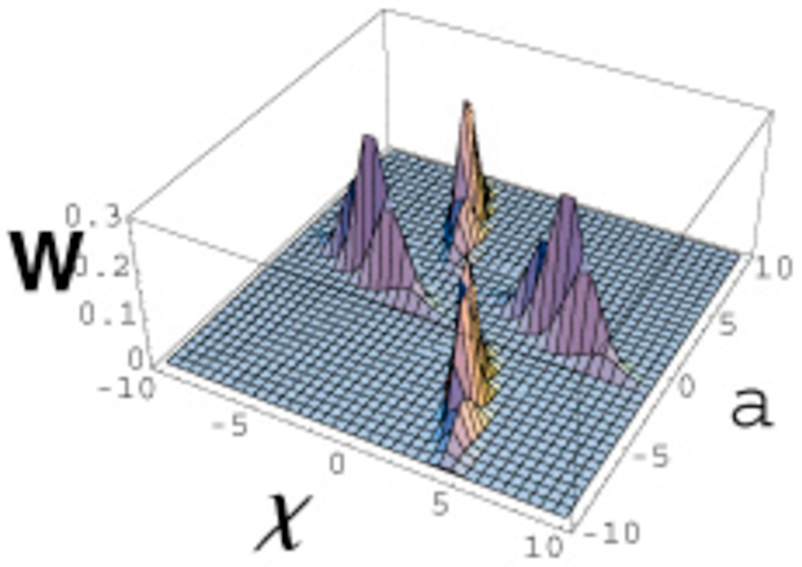}
\\
(a) \hspace{5cm} (b) \hspace{5cm} (c)
\caption{Plots of the Wigner functions of (a) a coherent state, $\beta=1$,
(b) a squeezed state, $\beta=2$ and (c) a squeezed state, $\beta=1/2$,
for $\varphi_0=-0.3$ in the conformal scalar model with $k=1$. In each figure,
the Wigner function at times $t=0$, $t=\pi/2$, $t=\pi$, and
$t=3\pi/2$ are shown at once.}
\label{fig5}
\end{figure}

As already seen from the analysis with cumulants, the behavior of the Gaussian wave
packet seems very simple and its behavior seems similar to, or consistent with,
the result from the analysis of the wave packet of Wheeler--DeWitt equation in the
conformal scalar model
\cite{Kiefer0,Kiefer2}.

\subsection{The wave function and the Wigner function of the conformal scalar model
with
$k=-1$}

The equivalent Hamiltonian of the conformal scalar model with $k=-1$ can be
expressed by that of an inverted harmonic oscillator, as (\ref{asham}) in
Sec.~\ref{sec3}.
A typical Gaussian wave
function in our model (with $k=-1$) can be written as
\begin{eqnarray}
\Psi(Q',t)&=&\frac{1}{\sqrt{\pi^{1/2}A(t)}}\exp\left[\frac{i
\{({Q'}^2+{Q'_0}^2)\cosh t-2 Q'_0Q'\}}{2\hbar
\sinh t}\right]\nonumber \\
& &\times\exp\left[-\frac{i\beta}{2\hbar\sinh t}\frac{(Q'-Q'_0\cosh t-P'_0\sinh
t)^2}{A(t)}\right]\,,
\end{eqnarray}
where
\begin{equation}
A(t)\equiv\beta\cosh t+i\frac{\hbar}{\beta}\sinh t\,,
\end{equation}
and $Q'_0$, $P'_0$, and $\beta$ are constants.
The probability density is given by
\begin{equation}
|\Psi(Q',t)|^2=\frac{1}{\sqrt{\pi}|A(t)|}
\exp\left[-\frac{(Q'-Q'_0\cosh t-P'_0\sinh t)^2}{|A(t)|^2}\right]\,.
\end{equation}
The initial wave packet gives $\langle Q'\rangle=Q'_0$ at $t=0$.
Note that the second order cumulant $\kappa_{2,0}$ is given by
$|A(t)|^2=\frac{\hbar}{2}\left(\frac{\beta^2}{\hbar}\cosh^2t+
\frac{\hbar}{\beta^2}\sinh^2t\right)$.
The initial condition on the coherence is given by the constant $\beta$
and $\beta=\sqrt{\hbar}$ gives a coherent state. In this case,
$|A(t)|^2=(\hbar/2)\cosh 2t$.


The Wigner function of the Gaussian wave packet in the model with $k=-1$ is given
by
\begin{eqnarray}
& &W(Q', P',t)\nonumber \\
& &=\frac{1}{\pi\hbar}
\exp\left[-\frac{1}{\beta^2}(Q'\cosh t-P'\sinh
t-Q'_0)^2-\frac{\beta^2}{\hbar^2}(P'\cosh t-Q'\sinh
t-P'_0)^2\right]\,.
\end{eqnarray}
Note that the Wigner function is also positive in this case.

\begin{figure}[ht]
\centering
\includegraphics[width=5cm]{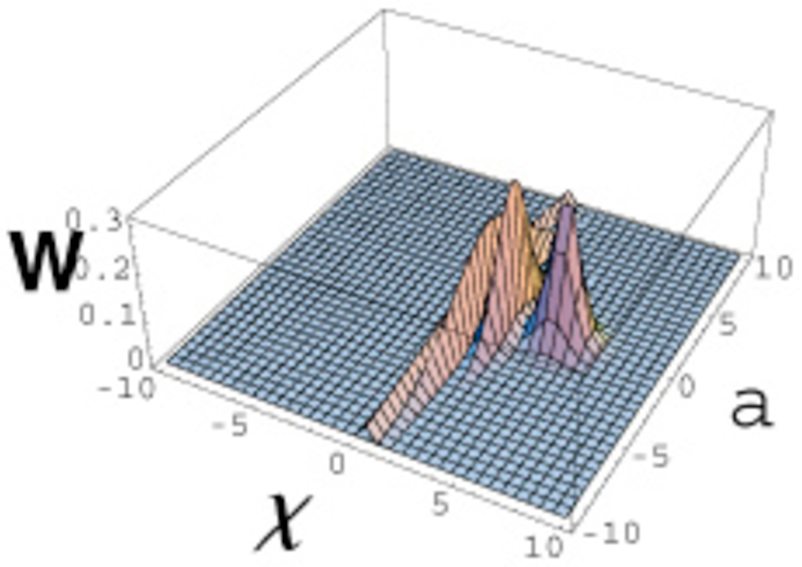}\quad
\includegraphics[width=5cm]{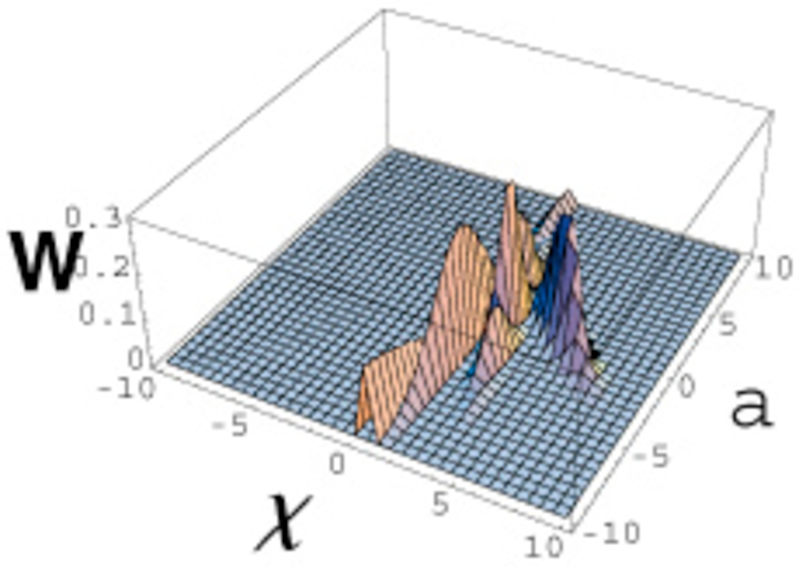}\quad
\includegraphics[width=5cm]{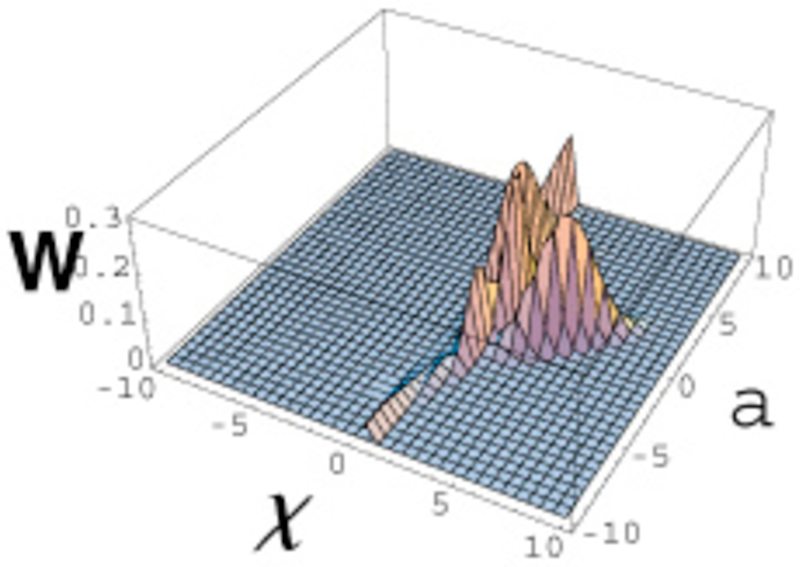}
\\
(a) \hspace{5cm} (b) \hspace{5cm} (c)
\caption{Plots of the Wigner functions of (a) a coherent state, $\beta=1$,
(b) a squeezed state, $\beta=2$ and (c) a squeezed state, $\beta=1/2$,
for $\varphi_0=0$ in the conformal scalar model with $k=-1$. In each figure, the
Wigner function at times
$t=0$,
$t=\pi/4$, $t=\pi/2$, and
$t=3\pi/4$ are shown at once.}
\label{fig6}
\end{figure}

\begin{figure}[ht]
\centering
\includegraphics[width=5cm]{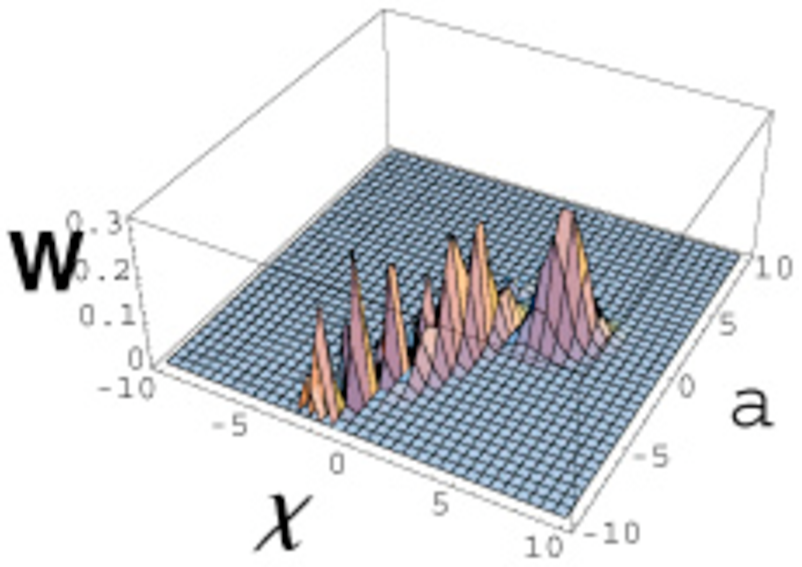}\quad
\includegraphics[width=5cm]{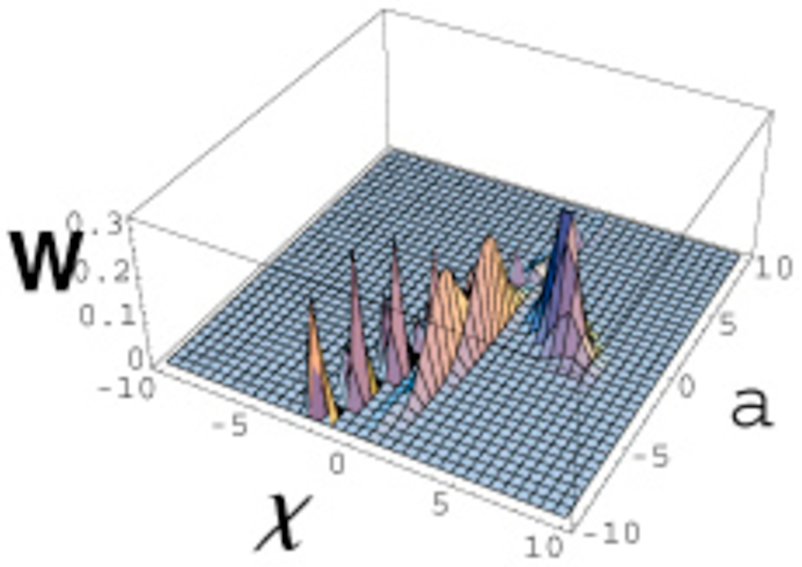}\quad
\includegraphics[width=5cm]{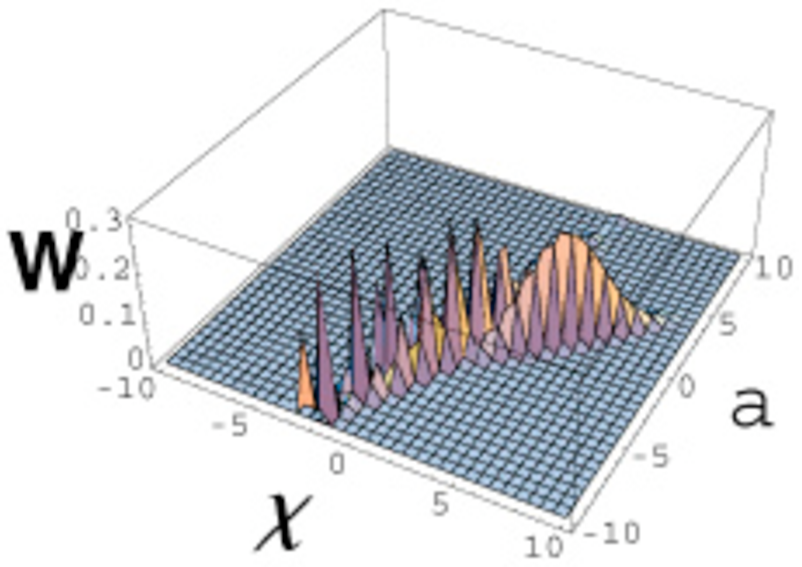}
\\
(a) \hspace{5cm} (b) \hspace{5cm} (c)
\caption{Plots of the Wigner functions of (a) a coherent state, $\beta=1$,
(b) a squeezed state, $\beta=2$ and (c) a squeezed state, $\beta=1/2$,
for $\varphi_0=0.3$ in the conformal scalar model with $k=-1$. In each figure, the
Wigner function at times $t=0$, $t=\pi/4$, $t=\pi/2$, and
$t=3\pi/4$ are shown at once.}
\label{fig7}
\end{figure}

\begin{figure}[ht]
\centering
\includegraphics[width=5cm]{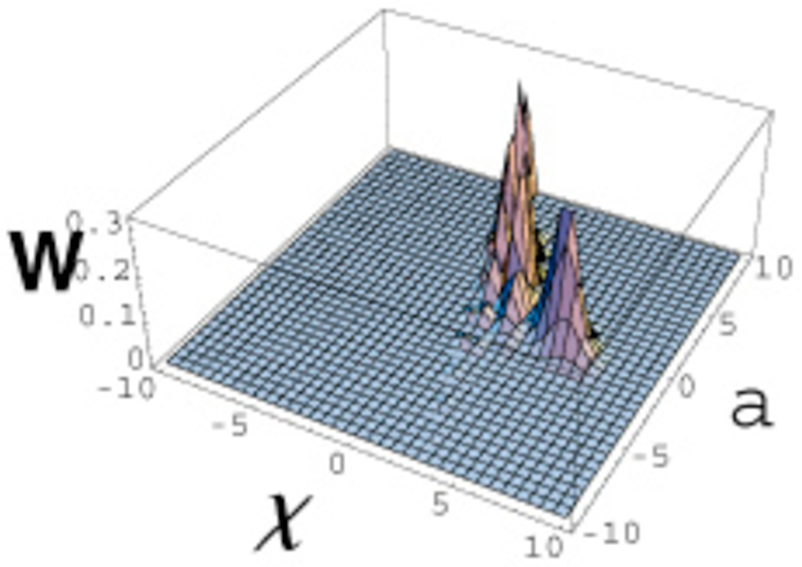}\quad
\includegraphics[width=5cm]{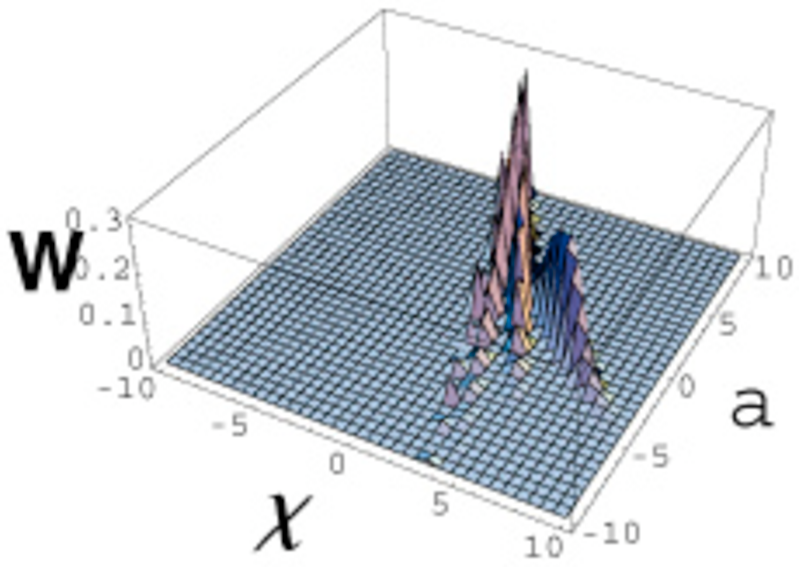}\quad
\includegraphics[width=5cm]{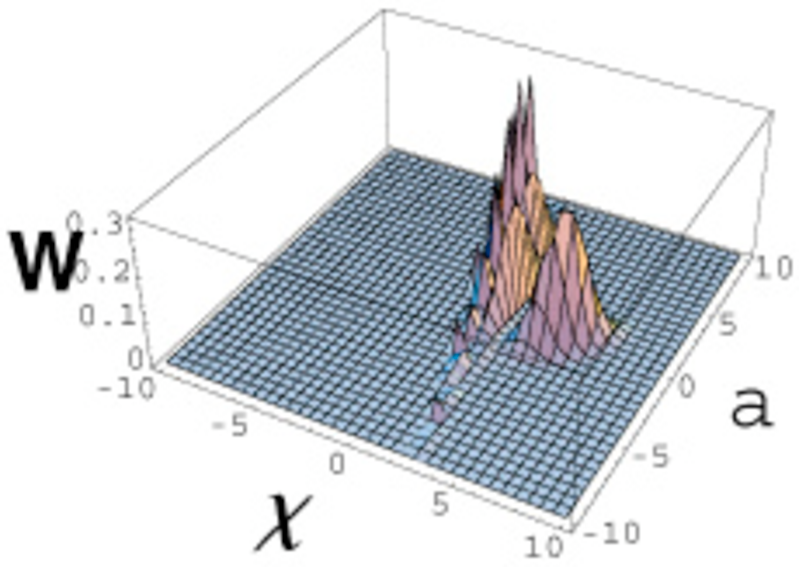}
\\
(a) \hspace{5cm} (b) \hspace{5cm} (c)
\caption{Plots of the Wigner functions of (a) a coherent state, $\beta=1$,
(b) a squeezed state, $\beta=2$ and (c) a squeezed state, $\beta=1/2$,
for $\varphi_0=-0.3$ in the conformal scalar model with $k=-1$. In each figure,
the Wigner function at times $t=0$, $t=\pi/4$, $t=\pi/2$, and
$t=3\pi/4$ are shown at once.}
\label{fig8}
\end{figure}

Remembering the relation $(Q', P')$ and $(a, \chi)$ (\ref{relations2})
in Sec.~\ref{sec3}, we can plot the Wigner function as a two-parameter function of
$a$ and $\chi$. We set $\hbar=1$, and $
a
(0)=0$ and $
\chi
(0)=5$ in the figures. Fig.~\ref{fig6} shows the Wigner
functions of (a) a coherent state,
$\beta=1$, (b) a squeezed state, $\beta=2$ and (c) a squeezed state, $\beta=1/2$,
for $\varphi_0=0$. 
Fig.~\ref{fig7} shows the Wigner functions of the same states
for $\varphi_0=0.3$. 
Fig.~\ref{fig8} shows the Wigner functions of the same states
for $\varphi_0=-0.3$. 
In each figure, the Wigner function at times $t=0$, $t=\pi/4$, $t=\pi/2$, and
$t=3\pi/4$ are shown at once.


The Wigner functions directly demonstrate the decoherence of the initial wave
packets in the model with $k=-1$.%
\footnote{As seeing from the classical solution, there is a bouncing solution,
where $a>0$, in the model with $k=-1$. We found, however, that the time scale of
the decoherence is the same as the classical evolution and further minute analysis
is needed for the solution. We leave the analysis for a separate study.}

\section{Conclusion}
\label{conclusion}

We have provided equivalent Hamiltonians for two integrable cosmological models,
in order to have Schr\"odinger-type equations. In the present paper, we have
investigated semiclassical behavior  of the models using quantum dynamics
incorporating cumulants. 
In the
Liouville scalar model with $U>0$, the value of $\kappa_{0,2}$ tends to take a
minus value and $\kappa_{2,0}$  diverges around $p\fallingdotseq -1$. Therefore, we
have to incorporate higher order cumulants to obtain precise quantum effect. On the
other hand, the evolution of the cumulants in the model with $U<0$ are more
moderate than the model with $U>0$.
In the conformal scalar model, the wave function has been
solved and we found that the Wigner function indicates the time-evolution of the
Gaussian wave packet in quantum cosmology.
The functions behave
as harmonic oscillators in the model with $k=+1$. In other words, the peaks of
$W(a,\chi,t)$ rotate on the $(a,\chi)$ plane with time evolution. Meanwhile, the
Wigner functions show unstable packets in the model with $k=-1$. So a
wave packet that is sufficiently localized is going to spread out and becomes
delocalized at a later time. Moreover, $\varphi_0$ changes the trajectories on
the $(a,\chi)$ plane. Additionally, if the Wigner functions with $\varphi_0\ne 0$
have a single peak at $t=0$, however, they split out to be multiple peaks with the
lapse of time.


The most questionable feature of the present approach would be the use of the
Hamiltonian not being bounded below in the equivalent system, though similar cases
have been studied in a number of papers to date
\cite{BK99a,BK99b,Conne,Sierra,ST,BBM,Barton}.
However, in light of general context of quantum cosmology, features such as
decoherence and delocalization seem to be possible properties \cite{Kiefer0}, so we
think it is meaningful to explore further formulation.


As a future study, we should examine the precise behavior of wave functions and the
Wigner functions in our models by various approximation method including simple WKB
method and by numerical calculations.
On the other hand, cosmological models with a cyclic coordinate in the
minisuperspace can be found in theory that possesses shift symmetries.
Generalization of the present analysis to other symmetric models will bring us
with a new insight into quantum cosmology.

We showed examples for restricted subclass of integrable models.
In the present study, we considered the dynamics in minisuperspace
and so we cannot exhibit the feature of all of the degree of freedom in the
original theory through the effective Hamiltonian (or Lagrangian). In lower
dimensions, however, it is known that the B\"acklund transformation enables us to
treat some integrable models as free field theories \cite{BCT}. In such a case,
quantum cosmology and the treatment of constraints may be studied exactly with the
equivalent Hamiltonian.  Therefore, one important subject to study is in lower
dimensional quantum cosmological models.

As a generic cosmological model, not so an academic theme, it is very important to
incorporate the contribution of matter and consideration of non-integrable models
is essential.  As we showed in the present paper, reduction of phase space is
considered to be an important key point in the constraint system in our analysis.
 We suppose that the recent
study of the Hamiltonian analysis of gravity \cite{RGP} may give a certain
direction in the investigation of quantum cosmology for this reason.
We consider that supersymmetric quantum cosmology \cite{Moniz}, which deals with an
extended constraint system, may be an interesting target to study about reduction
of phase space.

\appendix

\section{Canonical transform of the Liouville Hamiltonian}\label{AA}

This Appendix \ref{AA} reviews the canonical transform of the Liouville Hamiltonian
\cite{Ghandour}.
We consider the following Liouville Hamiltonian
\begin{equation}
h=\frac{1}{2}\pi^2-\frac{U}{2}e^{2\lambda x}\,,
\end{equation}
where $\pi$ is the conjugate momentum of the variable $x$.
Now, we consider a generating function
\begin{equation}
F(x,X)=\lambda^{-1}\sqrt{U}e^{\lambda x}\cosh\lambda X\,,\quad (U>0)
\end{equation}
\begin{equation}
F(x,X)=\lambda^{-1}\sqrt{|U|}e^{\lambda x}\sinh\lambda X\,,\quad (U<0)
\end{equation}

Since
\begin{equation}
\frac{\partial F}{\partial x}\dot{x}+\frac{\partial F}{\partial X}\dot{X}
=\frac{dF}{dt}\,,
\end{equation}
the canonical transformation $(x, \pi)\rightarrow(X, \Pi)$ gives
\begin{equation}
\pi=\frac{\partial F}{\partial x}=
\sqrt{U}e^{\lambda x}\cosh\lambda X\,,\quad
\Pi=-\frac{\partial F}{\partial X}=-\sqrt{U}e^{\lambda
x}\sinh\lambda X\,,\quad (U>0)
\end{equation}
\begin{equation}
\pi=\frac{\partial F}{\partial x}=
\sqrt{|U|}e^{\lambda x}\sinh\lambda X\,,\quad
\Pi=-\frac{\partial F}{\partial X}=-\sqrt{|U|}e^{\lambda
x}\cosh\lambda X\,,\quad (U<0)
\end{equation}

Then, we find
\begin{equation}
h=\frac{1}{2}\Pi^2\,.
\end{equation}
Note that
the constraint in the original $x$-$y$ system of the Liouville model requires
identification $\Pi\approx \pi_y$ and thus $X\approx y-y_0$, which reproduces the
first order equation in Sec.~\ref{sec2} at classical level.
Note also that $\bar{H}$ in Sec.~\ref{sec2} has the same form as $\Pi$ here.


\bibliographystyle{apsrev4-1}

\end{document}